# Cosmic Rays X. The cosmic ray knee and beyond: Diffusive acceleration at oblique shocks

A. Meli[1] and P. L. Biermann[2,3]

[1] Institut für Physik, University of Dortmund, Otto-Hahn-Str. 4, 44227 Dortmund, Germany
  e-mail: `meli@physik.uni-dortmund.de` *
[2] Max Planck Institut für Radioastronomie, Auf dem Hügel 69, 53121 Bonn, Germany
[3] Department of Physics and Astronomy, University of Bonn, Germany
  e-mail: `plbiermann@mpifr-bonn.mpg.de`

**Abstract.** Our purpose is to evaluate the rate of the maximum energy and the acceleration rate that cosmic rays acquire in the non-relativistic diffusive shock acceleration as it could apply during their lifetime in various astrophysical sites, where highly oblique shocks exist. We examine numerically (using Monte Carlo simulations) the effect of the diffusion coefficients on the energy gain and the acceleration rate, by testing the role between the obliquity of the magnetic field to the shock normal, and the significance of both perpendicular cross-field diffusion and parallel diffusion coefficients to the acceleration rate. We find (and justify previous analytical work - Jokipii 1987) that in highly oblique shocks the smaller the perpendicular diffusion gets compared to the parallel diffusion coefficient values, the greater the energy gain of the cosmic rays to be obtained. An explanation of the cosmic ray spectrum in high energies, between $10^{15}$eV and about $10^{18}$eV is claimed, as we estimate the upper limit of energy that cosmic rays could gain in plausible astrophysical regimes; interpreted by the scenario of cosmic rays which are injected by three different kind of sources, (a) supernovae which explode into the interstellar medium, (b) Red Supergiants, and (c) Wolf-Rayet stars, where the two latter explode into their pre-supernovae winds.

## 1. Introduction

It is been quite some time since the mechanism of particle shock acceleration (Fermi, 1949; Krymskii, 1977; Axford et al., 1978; Bell, 1978a, 1978b; Blandford & Ostriker, 1978) was proposed. It is believed to account for the origin of cosmic rays over the entire cosmic ray spectrum range detected. But questions still remain and not fully settled. The debate has reached a consensus about firstly, that *most* of the cosmic rays (energies up to around $10^{15}$eV) are produced in the shock waves for example, of Super Nova (SN) explosions (e.g. Ginzburg, 1966; Lagage & Cesarsky, 1983; Drury, 1983; Silberberg et al., 1990) into the interstellar medium, or into a stellar wind (e.g. Völk & Biermann 1988). Secondly, particle acceleration in Active Galactic Nuclei (AGN) hot spots/jets, Pulsar winds and Gamma Ray Bursts (GRBs), has been claimed to play an important role to the very high energy cosmic ray production. Furthermore, a few arguments exist that try to give a coherent view about cosmic rays (for a review see, Biermann et al., 2005). Some of those are:

*Send offprint requests to*: A. Meli
 * Present address: Institut für Theoretische Physik IV: Weltraum- und Astrophysik, Ruhr-Universität Bochum, 44780 Bochum, Germany

1. Explosions of stars into the interstellar medium provide most of the Hydrogen in cosmic rays, and most of the electrons, and little of the other elements. This contribution cuts off near a few hundred TeV.
2. Explosions of stars into their stellar winds provide everything else, all the heavier elements, even at low energies, starting with almost all of the Helium. Here the magnetic field is already large in stellar winds, or is enhanced by amplification in the shock as suggested by Bell, (2004) and Völk et al., (2005). These contributions reach 3 EeV.
3. A few nearby radiogalaxies may provide all cosmic rays above $3\,10^{18}$ eV energy per particle (e.g. Biermann & Strittmatter, 1987).

Generally speaking, the spectral region between knee and ankle is contested as to where it physically comes from, shocks in winds, shocks in bubbles, shocks in the interstellar medium (ISM), second order Fermi, galactic wind shocks, active binary systems, extragalactic, etc.

Furthermore on this subject, Biermann (1993) proposed that cosmic rays up to about $3 \times 10^{18}$eV arise from shocks which expand within the cavity of exploded stars. On the other hand in theoretical works of Biermann & Strittmatter (1987), Vietri (1995, 1998), Waxman (1995), and in numerical simu-lations of Meli & Quenby (2003a, 2003b) - which justified the former - has been proposed that cosmic rays beyond the ankle ($> 3 \times 10^{18}$eV), may originate from GRB fireballs and AGN hotspots where the plasma flow velocities and shock formations are above the non-relativistic speeds. The latter have also shown using Monte Carlo simulations, that when very large gamma flows apply (i.e. plasma flows having speeds of a substantial fraction of the speed of light), the obliquity of the magnetic field to the the shock normal does not affect the energy gain of the particles as they cross the shock. In that work it was considered that the parallel coefficient $\kappa_\parallel$, is much larger then the perpendicular one $\kappa_\perp$. In the non-relativistic limit, Jokipii (1987) has theoretically shown that the direction of the average magnetic field at the shock seems to have a large effect for the acceleration mechanism when the $\kappa_\parallel$ is much larger than the $\kappa_\perp$, and as it has particularly shown, the energy gain of the particles increases as $\kappa_\perp$ decreases.

In this paper we aim to test numerically, applying Monte Carlo simulations, the above theory concerning the particles' diffusion behaviour in *non-relativistic* highly oblique shocks and to investigate the rate of the energy gain in the diffusive shock acceleration by allowing both parallel and perpendicular coefficients in the acceleration mechanism as it would apply to stellar winds. We remind the reader that in a standard magnetic field topology in a stellar wind, the magnetic field is asymptotically almost perfectly perpendicular to the radial direction, so that by necessity any radial shock is very oblique.

The structure of this paper is as follows: First we proceed presenting the analytical expressions for the diffusion coefficients while mentioning the arguments for the rate of energy gain and maximum energy involving diffusion coefficients and suggested limits; we present briefly the numerical code used for our test. Finally the results are been presented and we conclude by making predictions about 'realistic' astrophysical environments.

## 2. Parallel and perpendicular diffusion coefficients and analytical limitations

Much work has been done over the years (e.g. Jokipii, 1966, 1967, 1972, 1987; Wentzel, 1974; Völk, 1975; Skilling, 1971, 1975; Webb, 1985, 1989; Kirk et al., 1988; Schlickeiser, 1989a, 1989b) concerning studies on the kinetic theory of charged particles propagating in turbulent electromagnetic fields. As an example, the diffusion coefficients were discussed in detail by Bell (1978a) and he noted that the shocks themselves generate turbulent motions, so that the region just behind the shock is highly turbulent. This was observed in the Earths's bow shock. One could consider the cosmic rays as a very hot *collisionless* plasma. In a collisionless plasma, collisions are defined where there is a resonant wave-particle encounter. When these stream through a shocked gas they may collectively generate hydromagnetic waves which *then* scatter the cosmic rays themselves. There can also be pre-existing turbulence due to a gas cloud or wind motion, prescribed by the specific site of the transport and acceleration. Initially, scattering of these cosmic rays requires fluctuations in the magnetic field. Scattering is due to real waves moving at Alfvén speed in the ionized gas,

$V_A = B(4\pi\rho_i)^{-1/2}$ and in the direction in which the cosmic rays diffuse if caused by particle beam instabilities, where $\rho_i = m_p n_i$ and $m_p =$ the mass of proton. A simple example is the following. If we assume that an ambient magnetic field is $B = 3 \times 10^{-6}$G and the density of both electrons and ions present in the plasma is $n_i = 3 \times 10^{-3}$cm$^{-3}$ (e.g. Snowden et al., 1997 ) then $V_A$ equals 100 km sec$^{-1}$.

The scattering has two critical consequences. Firstly, it limits the rate at which cosmic rays can escape an astrophysical site and this may account for the near-isotropy and the ages the cosmic rays have. Those cosmic ray traits can be given as well by the effective mirroring where the particles trapped in the confined plasma, can get reflected backwards and forwards on both sides of the shock. Secondly, a specific mechanism is provided by which energy and momentum is transferred by cosmic rays to the interstellar medium. Then it follows that also the interstellar gas around the astrophysical site is heated and accelerated by this process, thus one can claim that it is difficult in the first place for one to distinguish when the acceleration and the so called transport of the particles begins.

Regarding the motion of the particles in a turbulent electromagnetic field, Lindquist (1966) finds that the relativistic generalization of Boltzmann's equation leads to

$$\frac{dx}{d\tau}\frac{\partial F_o}{\partial x} + \frac{dp}{d\tau}\frac{\partial F_o}{\partial p} = \left(\frac{\partial F_o}{\partial \tau}\right)_{coll}, \qquad (1)$$

where, $F_o$ is the invariant phase space density, $\tau$ is the proper time and $(\frac{\partial F_o}{\partial \tau})_{coll}$ represents the rate of $F_o$ with respect to proper time due to collisions. A clear distinction is made between processes such as the particle's pitch angle scattering and the radiative reaction (both influence the particle's trajectory), which are to be considered as a part of the collision operator and the Lorentz force, exerted on the particle due to external electromagnetic fields. The above equation has been solved for many conditions (e.g. Toptygin, 1980). It can be shown that the time for acceleration from an initial momentum $p_o$ to a momentum $p_1$ may be written as

$$\tau_{acc} = \frac{3}{V_1 - V_2} \int_{p_o}^{p_1} \left(\frac{\kappa_1}{V_1} + \frac{\kappa_2}{V_2}\right)\frac{dp}{p}, \qquad (2)$$

where one can assume that $\kappa_1 = \kappa$ as a function of momentum $p$ upstream, and $\kappa_2 = \kappa$ in the downstream region of the shock. At the first instance we could understand that only the diffusion coefficient $\kappa$ and the velocity of the plasma are those parameters that determine the acceleration rate. Furthermore a few authors have shown (e.g. Jokipii, 1987) that the diffusion coefficient $\kappa$ may be given by the sum of the contribution of the diffusion coefficients parallel and perpendicular to the magnetic field ($\kappa_\parallel$ and $\kappa_\perp$), where $\psi$ is the angle between the magnetic field and a shock front placed on the $x$ axis, $\kappa = \kappa_\parallel \cos^2\psi + \kappa_\perp \sin^2\psi$. Over the past years many theoretical and experimental studies from various authors (e.g. Jokipii, 1967 and Moussas et al., 1982a, 1982b, 1982c) have not yet succeeded to clarify the expected values of $\kappa_\parallel$ or $\kappa_\perp$ in given astrophysical circumstances. It seems though up to now, that the value of the parallel diffusion coefficient ($\kappa_\parallel$) tends to be substantially higher (at least from interplanetary measurements) than the perpendicular diffusion coefficient ($\kappa_\perp$) but, $\kappa\perp$ may approach $\kappa_\parallel$ in some situations.

In a theoretical work Jokipii (1987) investigated the rate of the energy gain and the maximum energy in non-relativistic shocks, which can be attained in given conditions, such as the effect of a highly oblique magnetic field and the $\kappa$ to the scattering of the particles across a shock front. Briefly, he showed that *if the perpendicular diffusion is much smaller then the parallel one, particles can gain considerable energy in quasi-perpendicular shocks compared to quasi-parallel ones* and he estimated that some *limits* may arise. He attempted to estimate constraints between the mean free path value of the scattered particle across the magnetic field lines in relation to both parallel and perpendicular diffusion coefficients and the energy that a particle will achieve. He has generally shown that for the standard kinetic relationship between perpendicular and parallel diffusion the following condition should hold

$$\kappa > r_g V_{sh}, \tag{3}$$

where $V_{sh}$ is the velocity of the shock, or else the following must satisfied

$$\frac{\lambda_\parallel}{r_g} < \frac{\upsilon}{V_{sh}}, \tag{4}$$

where $\upsilon$ is the velocity of the particle. Jokipii argues that the above relations should hold, taking under consideration the *standard kinetic theory* approach. On the other hand, for the quasilinear theory where the contribution of the field random walk or meandering could play a significant role (Barge et al., 1984), the above relations will change and the particle acceleration behaviour will be affected accordingly.

Following the above analysis, what we do is to conduct different Monte Carlo simulation runs by violating or holding the inequality (given by relation 4) and try to obtain, - by using different inclinations of the magnetic field with respect to the shock normal - the average gain of a particle in a complete cycle at the shock frame. We use a range of angles, between the quasi-parallel and up to perpendicular ones. The velocity of the upstream fluid is much less than a considerable fraction of light, thus the condition allows us to use a long range of angles without being 'trapped' in the limits of a superluminal shock configuration, where a transformation to the de Hoffmann-Teller frame (de Hoffmann & Teller, 1950) is not possible, allowing that $V_{HT} < V_{sh} \tan \psi$. We remind the reader that the de Hoffmann-Teller frame is a reference frame where the magnetic field and the velocity vector of the flow are parallel upstream and downstream with the electric field, $E = 0$. Our tests show (see following figures) that when $\kappa_\parallel \gg \kappa_\perp$ and the shock is almost perpendicular ($\psi_B \to 90^o$) then the particle gains faster much higher energies than when the magnetic field is almost parallel to the shock normal ($\psi_B \to 0^o$), confirming the analytical solutions. It is straightforward from the calculations, that the acceleration rate of the particles under the above favoured conditions is smaller in the near perpendicular shock configurations than in the parallel ones and the limits and applications will be described in section 4.

## 3. Numerical simulations

A Monte Carlo code was constructed in order to simulate non-relativistic near parallel and oblique shocks with $0^o \lesssim \psi \lesssim 90^o$ where $\psi$ is the angle between the shock normal and the magnetic field, seen in the shock frame. The velocities of the upstream plasma flow are kept between $0.01c$ and $0.05c$, which correspond to astrophysical environments with non-relativistic shocks. We note that SN shock speeds can reach up to $0.05c$. For the description of the physical quantities necessary throughout the simulations we use the local fluid frame of reference (1,2 for upstream and downstream respectively), the normal shock frame (sh) and the de Hoffman-Teller frame (HT). We isotropically inject particles from upstream with $\upsilon \approx c$, towards the shock where are allowed to scatter in the fluid frame and across the shock discontinuity. We follow each particle trajectory across the shock according to the standard theory 'jump conditions' and each particle leaves the system when it 'escapes' far downstream at a spatial boundary or if it reaches a well defined maximum energy $E_{max}$. For the non-relativistic shocks the compression ratio, $r = V_1/V_2$, where $V_1$ and $V_2$ are the upstream and downstream plasma flow velocities, is kept equal to 4. We note here that in the test-particle theory of diffusive shock acceleration, the spectral index $\sigma = 3r/(r-1)$ depends on the compression ratio $r$. For strong shocks in an ideal gas, ($c_p/c_v = 5/3$) which gives $\sigma = 4$ (Axford et al., 1978; Krymskii, 1977; Blandford & Ostriker, 1978).

A splitting technique is used similar to one used in Monte Carlo simulations of Meli & Quenby (2003a, 2003b), so that when an energy is reached such that only a few accelerated particles remain, each particle is replaced by a number of $N$ particles of statistical weight $1/N$ so as to keep roughly the same number of particles being followed. For the diffusion along the pressumed magnetic field lines, a guiding centre approximation is used, where the particle trajectory is followed in two-dimensional, pitch angle and distance along **B** space. The mean free path is calculated in the respective fluid frames by the formula: $\lambda = \lambda_\circ p_{1,2}$, assuming a momentum dependence to this mean free path for scattering along and across the field and related to the effective diffusion coefficient. During the simulation the particle is allowed to cross the assumed field lines. As the particle is injected upstream an initial $\theta$ and $\phi$ is given to it which are randomized between each scattering, as to calculate the position of the particle in each step following the relations $x_o = r_g s \sin \psi, y_o = -r_g s \cos \psi$ where $s$ is given by, $s = \sin \theta_i \sin \phi_i - \sin \theta_{i-1} \sin \phi_{i-1}$. The probability *Prob*, that a particle will move a distance $\Delta s$ along the field lines at pitch angle $\theta$ before it scatters, is given by the expression $Prob(\Delta s) \sim exp(-\Delta s/\lambda)$ and the probability of finding a particle in a pitch angle interval $d\mu$ where $\mu = cos\theta$ is $\mu$, and the time between collisions is $\upsilon|\mu|$ for particle velocity $\upsilon$ and $\mu$ chosen randomly between $-1$ and $+1$. For further discussion on the scattering the reader is referred to the review of Quenby and Meli (2004). Furthermore, in the shock rest frame, the flow velocities ($V_1, V_2$) for upstream and downstream respectively, are parallel to the shock normal and the magnetic fields $\mathbf{B}_1$ and $\mathbf{B}_2$ are at an angle $\psi_1$ and $\psi_2$ to the shock normal respectively. We adopt a geometry with $x$ in the flow direction, positive downstream, $\mathbf{B}_1, \mathbf{B}_2$ in the $x - y$ plane and directed in the negative $x$ and $y$ directions and only $\mathbf{E}_z$ finite and in the positive $z$ direction. A Lorentz transformation is performed to the local fluid frame (1 or 2) each time the particle scatters across the shock. For the Lorentz transformations between the shock frame and the de Hoffmann-Teller frame, we need to boost by a $V_{HT}$ speed along the shock frame where, $V_{HT}$ is equal to $V_{sh} \tan \psi_1$. We will note that the transformation to the de Hoffmann-Teller frame though, is only possible if $V_{HT}$ is less the speed of light. This means that $\tan \psi_1 \leq 1$, which is actually in the limit (subluminal) we keep our calculations.

A number of $\sim 10^4$ particles of a weight equal to 1.0, are injected far upstream at a constant energy (e.g. $\gamma = 5$ -assuming particles are already relativistic in the wind). The particles are left to move towards the shock, they collide along the way with the scattering centers and consequently as they keep scattering between the upstream and downstream regions of the shock (its width is much smaller than the particle's gyroradius) they gain a considerable amount of energy. As we see the oblique shock, from the de Hofmann-Teller frame where the electric field is zero, a particle trajectory is prescribed by five coordinates and, the energy is conserved but the direction of the velocity vector changes. From these five quantities the magnetic moment ($p_M$) is conserved as the particle encounters the shock and as it scatters. Specifically, Hudson (1965) showed analytically, for non-relativistic flows and super-luminal case field inclinations, that the moment is conserved. Terasawa (1979) computed trajectories in this non-relativistic limit and demonstrated both near gyro isotropy and near conservation. Drury (1983), in a non-relativistic approximation ($V \ll c$) uses Liouville and gyrophase invariance to show conservation in both sub and super-luminal cases. So, if we denote the component of the particle momentum perpendicular to the magnetic field $\mathbf{B}$, as $p_\perp$, then the magnetic moment will be given by, $p_\perp^2/B$. In the de Hofmann-Teller frame $p = |\mathbf{p}|$ is also conserved, so we can find the downstream pitch angle which is given by $\mu_2 = \sqrt{(\mu_1^2 - \mu_{1,c}^2)/(1 - \mu_{1,c}^2)} \cdot (\mu_1/|\mu_1|)$ where, $\mu$ is the cosine of the pitch angle '1' and '2' denote the upstream and downstream quantities respectively. Furthermore, the theory for the oblique shocks postulates that from the conservation of the first adiabatic invariant one can find the new pitch angle in the downstream frame and similar transformations allow the particle scattering to be followed in this frame. The particle that crosses the shock from upstream, is transmitted only if its pitch angle is less than the critical pitch angle (cosine of the 'loss cone' angle), $\mu_c = \sqrt{\frac{\mathbf{B}_{HT,1}}{\mathbf{B}_{HT,2}}}$. Following the above relations of the conservation of the first adiabatic invariant, we can find the new pitch angle in the downstream frame and similar transformations allow the particle scattering to be followed in this frame.

## 4. Results

In this section results are presented based on a series of Monte Carlo simulations for particle shocks acceleration, while using a variety of different values for the diffusion coefficients $\kappa_\parallel$ and $\kappa_\perp$, and the angle $\psi$ between the magnetic field lines and the shock normal. In other words the aim is to test analytical assumptions and conclusions focusing on Jokipii's work (1987) regarding non-relativistic diffusive particle shock acceleration in highly oblique shocks. The behaviour of the particles seems most crucial, affecting the acceleration mechanism, by crossing a highly oblique shock while scattering along and across the magnetic field lines, by satisfying the inequality $\lambda/r_g < c/V_{sh}$ ($v \sim c$ is the particle's velocity). At the same time we measure the actual time that the particles need to get to high energies (e.g. $\gamma = 10^8$ corresponding to around $9 \times 10^{17}$eV for protons) and compare it with the standard theoretical expression for the acceleration time (see equation 2 and figure 1). The first point we would like to stress out here is that the results show that the high obliquity of the shock to the magnetic field lines could play the most crucial role for the mean energy gain of the diffusing particles confined within an appropriate simulation time scale and maximum momentum limit. Past works (e.g. Ostrowski, 1988; Takahara and Terasawa, 1991; Ellison et al., 1995) studied the obliquity of the shock dependent on the $E_{max}$ and found as well that there is more rapid acceleration in nearly perpendicular shocks than in parallel ones. Ostrowski argues that the assumption that the amplitude of the magnetic field perturbations is small, must be fulfilled for the approach used to be valid. Of course this assumption allows us to make the approximation that the magnetic moment of the particle is conserved while it interacts with the magnetic field discontinuity, an assumption that we follow in our simulations.

A typical run and the energy gain of a particle accelerated, by getting reflected off or transmitted across the shock surface (prescribed in - dimensionless - gammas) is shown in table 1. We note again that in the de-HT frame the intersection between the magnetic field and the shock has a velocity, $V_{sh} \tan \psi$ less or equal (for the almost perpendicular case) than $c = 1$. A particle hits the shock with a pitch angle which is lying inside or outside an 'angle loss cone' of $60^o$, following the standard theory of the conservation of the first adiabatic invariant at the de-HT frame, and after some interactions with the shock interface (reflections) will eventually be transmitted across it. Our simulations show that a number less than $\sim 20\%$ of the particles, reflects as many times as to get such high energies as $10^{18}$eV within prescribed momentum and space limits of the simulation box, while the rest leave the simulation earlier with much less energy. This trait is a key feature in understanding the momentum gain and distribution (see plots in figure 2). Particularly, in the top plot the 'classical' expected particle shock acceleration distribution momenta is shown. On the contrary, in the bottom plot of figure 2 we can see how the distribution of the momenta changes under the extreme cases of the nearly perpendicular shocks as we describe in section 2. We can observe how the distribution of the momentum gain of the accelerated particles extends into a tail, a trend that is clearly connected to the Lorentz transformation of the de Hoffmann-Teller frame. Further to the traits of the table 1, it is possible that the probability of the particle to lying within or outside the 'angle loss cone' described in section 3, the effect of $\kappa$, and the actual limitations that it is given to its values by following the analytical claims of Jokipii (1987), beam the particles with an 'enhanced probability' bouncing off the shock surface with an angle less than $60^o$, which determines the reflection condition of the diffusive particle.

In figure 1 we show the ratio of the simulation acceleration rate to the theoretical formula (see equation 2), versus $\lambda/r_g$, where $\kappa > r_g V_{sh}$ or in other words $\lambda/r_g < v/V_{sh}$. As we mentioned in section 2, in equation (2) $\kappa$ is given by: $\kappa = \kappa_\parallel \cos^2 \psi + \kappa_\perp \sin^2 \psi$, while we keep $\frac{\kappa_\parallel}{\kappa_\perp} = 1 + (\lambda/r_g)^2$. For the simulations we use plasma flow velocity equal to 0.01c where obviously the maximum Jokipii limit is at $\lambda/r_g = 100$ and for plasma flow velocity equal to 0.05c the maximum Jokipii limit is placed at $\lambda/r_g = 20$. Analytically equation (2) is valid in Jokipii's case, but has to be used with the coefficient parallel to the shock normal, which in the limit of a highly oblique shock is almost perpendicular to the magnetic field. A condition that it is quantitatively underpinned by Biermann (1993). It is found that the 'Jokipii limit' is faster by $c/3V_{sh}$ and the momentum boundaries for these runs were fixed at $10^8$ gamma (i.e. $\sim 9 \times 10^{17}$eV for protons). It is easily understood that we recover the standard expression, and yet the standard derivation fails, in terms of the first order Fermi acceleration picture (with all the steps in momentum small) for the nearly perpendicular shock case prescribed under the above conditions.

In figure 3 we show the log of the mean energy of the particles during the acceleration process versus the number of shock crossings for $V_{sh} = 0.01c$ and $V_{sh} = 0.05c$ (top and bottom plot respectively). We note that three shock crossings is a shock-cycle. What it is shown here is similar to figure 1 of Jokipii (1987). That is, the energy gain of the particles is increasing with the increasing angle between the upstream magnetic field and the shock normal, showing the same trend of the shock efficiency depending on its obliquity to accelerate the particles. The scattering mean free path is kept at $\lambda = 100 r_g$ and $\lambda = 20 r_g$ respectively, following the trend of Jokipii's limits for testing. Different magnetic field inclinations are shown, starting from $5^o, 25^o, 65^o, 87^o$ and $89^o$. We see the great difference in the energy gain of the particles for the almost perpendicular case compared to smaller shock inclinations. The mean energy is the calculated energy gained by the particles (taking the sum and dividing it with the number of particles times their statistical weight) immediately after interacting with the shock interface. Looking closely, one may see that in a shock-magnetic field orientation of around less than $65^o$ the particles do not gain any considerable energy, but for highly oblique shock orientations (i.e. $> 65^o$) there is a considerable particle energy gain in a few cycles, an effect that firstly stressed by Jokipii (1987). Plasma flow velocities of $0.01c$ and $0.05c$ were used here and it is obvious that only when $\psi \rightarrow 90^o$ and $\kappa_\parallel \gg \kappa_\perp$, there is a large energy again achieved after a few shock cycles.

Furthermore, in the plots of figure 4 we show the differential energy and time spectra for a simulation run using $V_{sh}=0.05c$ and $\psi = 85^o$, respectively. Note that in top plot the spectrum is not changed while just two points are modified: 1) Getting to the maximum energy in a distribution of large and small jumps in momentum (see table 1) and 2) the speed of the acceleration process (obviously affected by the limits) with which we reach the spectrum (see figure 1).

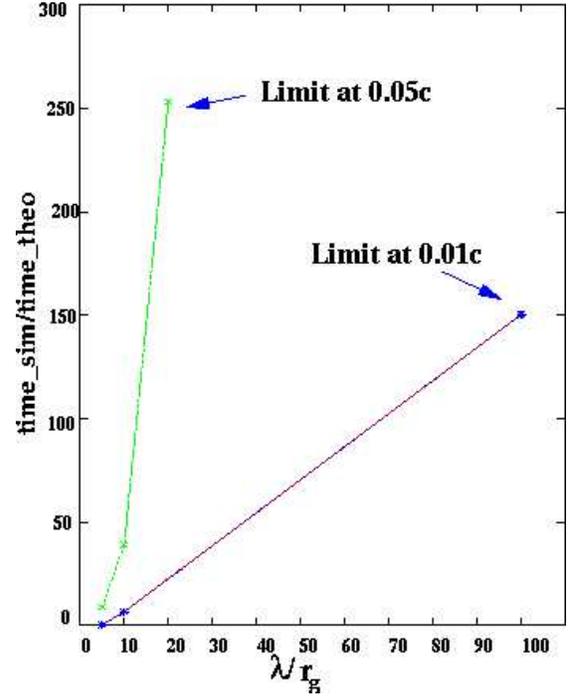

**Fig. 1.** Here we show the ratio of the simulation acceleration rate (time_sim) to the theoretical formula (time_theo) (see equation 2), versus $\lambda/r_g$, where $\kappa > r_g V_{sh}$ or in other words $\lambda/r_g < v/V_{sh}$. In equation (2) $\kappa$ is given by: $\kappa = \kappa_\parallel \cos^2 \psi + \kappa_\perp \sin^2 \psi$, while we keep $\frac{\kappa_\parallel}{\kappa_\perp} = 1 + (\lambda/r_g)^2$. For the simulations we use plasma flow velocity equal to 0.01c where obviously the maximum Jokipii limit is at $\lambda/r_g = 100$ and for plasma flow velocity equal to 0.05c the maximum Jokipii limit is placed at $\lambda/r_g = 20$. Analytically equation (2) is valid in Jokipii's case, but has to be used with the coefficient parallel to the shock normal, which in the limit of a highly oblique shock is almost perpendicular to the magnetic field. A condition that it is quantitatively underpinned by Biermann (1993). It is found that the 'Jokipii limit' is faster by $c/3V_{sh}$ and the momentum boundaries for these runs were fixed at $10^8$ gamma (i.e. $\sim 9 \times 10^{17}$eV for protons). As a numerical example here, we see that the acceleration rates almost converge together while the Jokipii limit should be lowered by at least a factor of 30 and converge with the standard analytical limitation of $\kappa = (1/3) r_g$.

### 4.1. Supernova shocks in stellar winds - An application

In 1958, Parker (1958) proposed the magnetic stellar wind model where the magnetic field lines of the wind plasma are initially in a radial direction. Biermann (e.g. 1993) proposed the model of spherical shocks in stellar winds which compresses an embedded tangential magnetic field which decreases with radius $r$ as $1/r$ and with colatitude $\theta$ as $\sin \theta$ by the full compression ratio of 4, and proposes that 'a principle of the smallest dominant scale' (Prandtl, 1925 and Karman & Howarth, 1938) (in real space or in velocity space) allows one to determine the relevant transport coefficients, which describe the overall transport of the particles in the magnetic field present in the shocked region.

In more detail in this model he does not invoke resonant scattering, but fast convective turbulence as the dominant process in the acceleration region. On the other hand the initial assumptions made in those works do not ignore resonant scat-tering, but use Jokipii's (1987) arguments to support the whole proposed model of the acceleration, supporting the idea of a 'permissible' scattering coefficient in oblique shock configurations.

| Gamma | Upstream Side (de H-T Frame) | Upstream Side (Shock Frame) | Upstream Fluid | Downstream Side (de H-T Frame) |
|---|---|---|---|---|
| inject upstream | | 0.69 | | |
| Reflected | 1.21 | 1.14 | 1.14 | |
| Reflected | 1.67 | 1.45 | 1.44 | |
| Reflected | 2.07 | 2.93 | 2.92 | |
| Reflected | 2.42 | 2.32 | 2.32 | |
| To downstream | | | | 2.65 |
| To upstream | 2.83 | 2.81 | 2.80 | |
| Reflected | 3.05 | 2.90 | 2.89 | |
| Reflected | 3.53 | 3.31 | 3.30 | |
| Reflected | 3.98 | 3.78 | 3.78 | |
| Reflected | 4.44 | 4.22 | 4.22 | |
| Reflected | 4.86 | 4.69 | 4.69 | |
| Reflected | 5.24 | 5.10 | 5.10 | |
| Reflected | 5.60 | 5.48 | 5.47 | |
| Reflected | 5.78 | 5.84 | 5.83 | |
| To downstream | | | | 6.19 |

**Table 1.** The Lorentz gamma factors gain for a particle in a typical simulation run for $V_{sh} = 0.05c$ and an inclination angle of $87^o$. The acceleration of a particle and the logarithm of its energy gain in different frames of reference is shown while it scatters and reflects off the shock surface until it exits the system of the acceleration. A shock-cycle is a a 3-time crossing of the shock. We mention that it is shown that a reflection is counted (in particle's energy gain) as cycle (Drury, 1983). We note that the high energy particle gains are due to the de Hoffmann-Teller frame Lorentz transformation where the shock behaves as if it was relativistic. The particle is injected upstream at $20 \times \lambda$ with a gamma of 5. The injection does not of course depend on number 20, as long as it is large compared to unity. The particle gets reflected by the shock interface four times, it crosses downstream, it crosses upstream, it gets reflected by the shock interface some times more, it crosses downstream, it leaves the system. The "particle's fate" of crossing or reflecting is calculated by the equations given in section 3, after its directions are properly randomized using random number generators for its scattering. As on sees, only in the nearly perpendicular shock configurations we have such high energy gains.

The normal argument (for relativistic particles) is that $\kappa = \frac{\lambda(E)c}{3} > \frac{r_g c}{3}$ where $\lambda(E)$ is the mean free path for resonant scattering of a particle of energy E and Larmor radius $r_g$. So, as we have already mentioned in a previous section the Jokipii 'limit' claims that possible values of the diffusion coefficient have to be larger than the gyroradius of the particle multiplied by the shock speed. But this specific condition is fulfilled for Biermann's (1993) approach only by a factor of $1 - V_2/V_1 < 1$, and it is close to unity only at the maximal energy, since in his model the shock speed and the radial scale of the system under consideration, give both the largest gyroradius as well as the diffusion coefficient. In the first instance we may say that it is clearer that by using equation (4), instead of $(1/3)(r_g c)$, one could claim that Jokipii is faster by $c/3V_{sh}$, and on the other hand simulations shows us that this factor should be even less in order to describe realistic maximum energy gain boundaries where a cut-off in the spectrum could be reached.

So, we claim here that a maximum cutoff in the spectrum could be $\sim 10^{17} eV$ (protons) or around $3 \times 10^{18}$ eV for iron. Again being in a realistic picture, we consider the general case of a shock expanding into a wind. Let is start from the mass conservation and allowing that all the mass of the gas in the wind is snow-plowed together and that the shock-shell conserves energy we have

$$\int_{r_\star}^{r_{sh}} 4\pi r^2 \frac{\dot{M}}{4\pi r^2 V_{wind}} dr = \Delta M = \text{mass in wind shock shell} \quad (5)$$

where $\Delta M = \frac{\dot{M} V_{wind}}{r_{sh}}$. We also have energy conservation and so, $(\frac{\dot{M} V_{wind}}{r_{sh}} + M_{sh,SN}) V_{sh}^2 = E_{SN}$. One could make the simple question if the piston $M_{sh,SN}$ is larger, or the wind-shell $\dot{M} V_{wind}/r_{sh}$. Testing with realistic numbers one find that the shock-shell is a small fraction of 1 solar mass, and so the piston always wins. This ensures that the shock velocity is exactly constant throughout the expansion of the shock through the wind. It follows, that the radius of the shock $r_{sh}$ and time are proportional to each other $r_{sh} = V_{sh} t$. In the shock frame now we identify $V_{sh}$ with $V_1$. The acceleration rate is given also by the equation $(\frac{dp}{dt})_{acc} = \frac{xp}{\tau_{acc}}$ with, $\tau_{acc} = 3(\frac{\kappa_1}{V_1} + \frac{\kappa_2}{V_2}) \frac{1}{V_1 - V_2}$ from Biermann (1993) $\kappa \propto \Delta r \Delta V$ and $x = 2.25$ below the knee, and $x = 1.729$ above the knee where $V_1$ is the upstream velocity, $V_2$ is the downstream velocity and $p$ is the trans-relativistic momentum.

The adiabatic losses (Völk & Biermann, 1988) are $(\frac{dp}{dt})_{ad} = \frac{p}{3} \frac{2}{r} V_1 \frac{V_1/V_2 - 1}{V_1/V_2}$. These two terms can be combined as $(\frac{dp}{dt})_{tot} = \frac{xp}{2t} - \frac{p}{2t} = (x-1)\frac{p}{t}$, with the solution $p = p_0 (t/t_0)^{x-1}$. Therefore one may have

$$p = p_0 (t/t_0)^{1.25}, \quad (6)$$

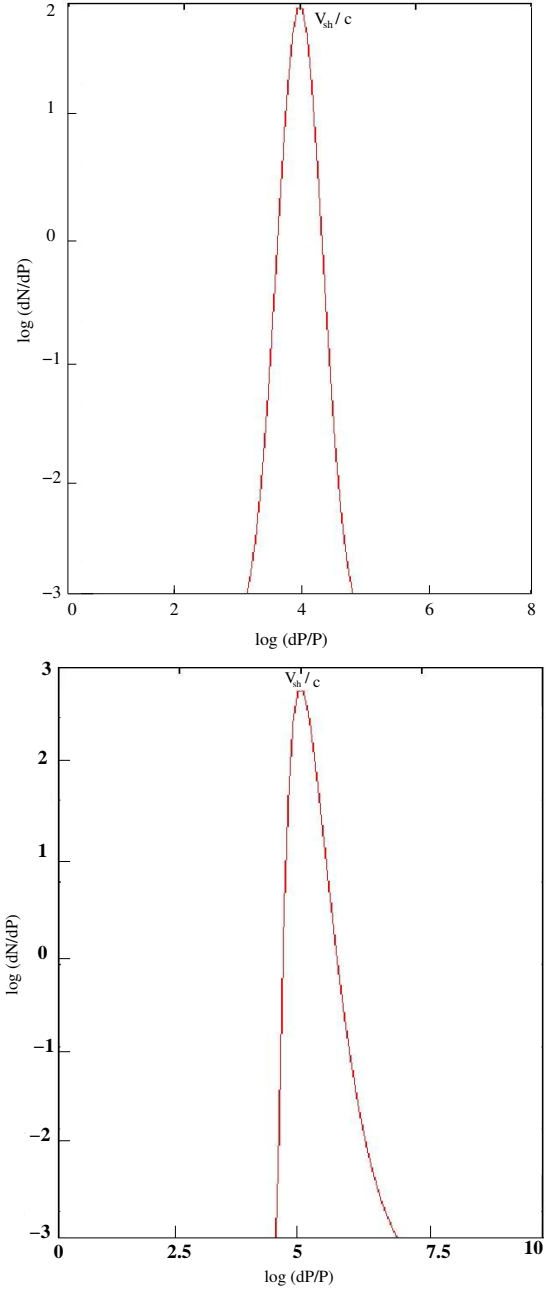

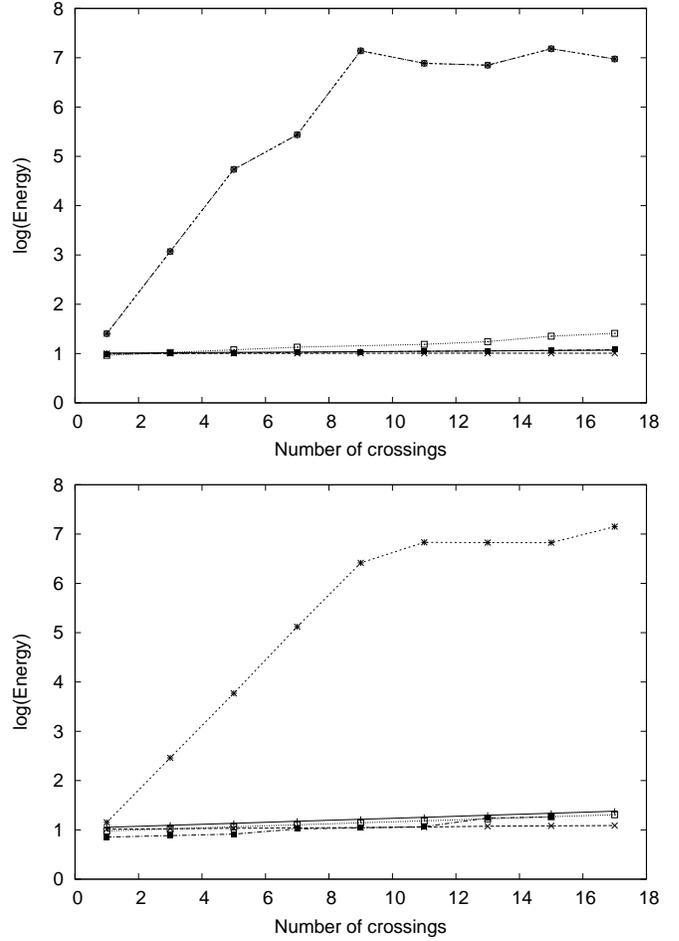

**Fig. 3.** The log of the mean energy of the particles versus the number of shock crossings for $V_{sh} = 0.01c$ (top plot) and $V_{sh} = 0.05c$ (bottom plot). Starting from the bottom for $5^o, 25^o, 65^o, 89^o$ (top plot) and $87^o$ (bottom plot) magnetic field inclination, respectively. We see the difference in the energy gain of the particles for the almost perpendicular case compared to smaller shock inclinations. An effect that Jokipii (1987) pointed out as well. Mean energy is the calculated energy gained by the particles (taking the sum and dividing it with the number of particles times their statistical weight) immediately after interacting with the shock interface (i.e. three shock crossings is a 'shock-cycle').

**Fig. 2.** Top plot: The 'classical' acceleration mechanism of the distribution of momenta of the accelerated particles for parallel and oblique shocks, will look as above. Bottom plot: In the extreme cases though of nearly perpendicular shocks (this study), the distribution for the momentum gain of the accelerated particles (expressed in Lorentz factor) will extend into a tail, in contrary to the top plot. This trend is clearly connected to the Lorentz transformation into the de Hoffmann-Teller frame.

up to the knee energy, and beyond

$$p = p_0 \left(t/t_0\right)^{0.729}.\qquad(7)$$

Since we start with $p_{inj}c = 3 \times V_{sh}m_pc = 0.15 m_pc^2$ and the knee is about $10^6\, m_pc^2$, the first phase corresponds to about a factor of $3 \times 10^5$ in radius and time, and a factor of 500 between radius and time from knee to ankle. Taking then a final radius for WR stars of $10^{19}$ cm, this implies, that the knee is established at a distance of $2 \times 10^{17}$ cm, and injection starts at $7 \times 10^{11}$ cm. Moreover, these numbers could be refined, with a specific model for the magnetic field topology and wind velocity in stellar winds of RSG and WR stars.

Here we are including adiabatic losses as well and one may see that even if we have more time, then the maximum energy cannot change anymore, since we have reached the maximum allowed by the spatial condition, fitting the Larmor motion into the available space. The other scenario is of course that if we have less time or space, then the maximum energy is not reachable anyway. Then of course the cutoff in the energy will be much lower than the proposed $\sim 10^{17}$ eV. The space in a stellar wind is radius/4, so $r/4$, from the Rankine-Hugoniot conditions for strong shocks for a gas with adiabatic index 5/3. And since the dominant magnetic field is tangential, the Larmor limit is a constant, as then $B(r) \times r$ is constant.

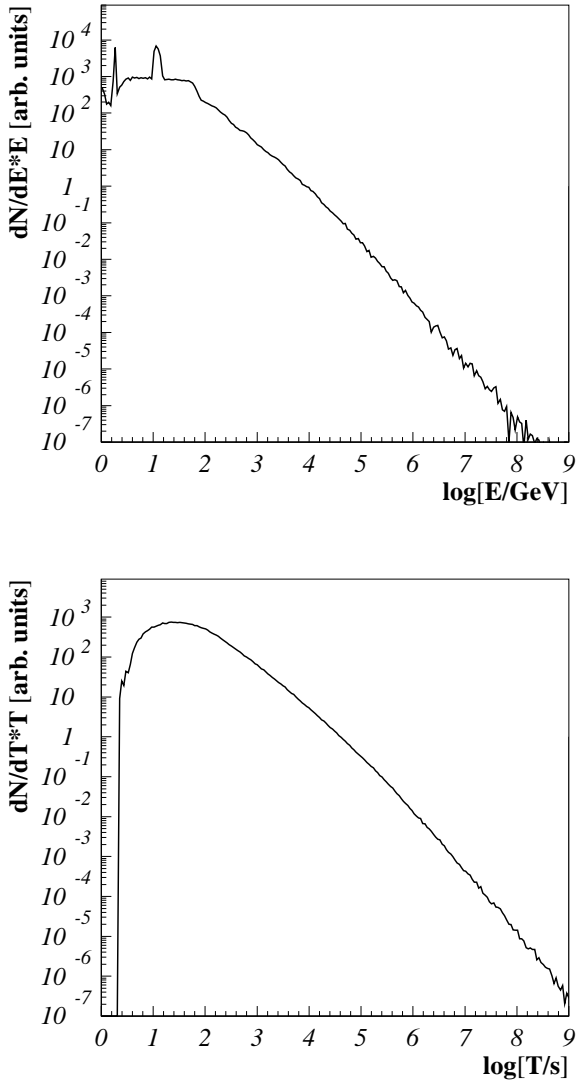

**Fig. 4.** Top plot: A typical spectrum for $85^o$ for $V_{sh}$ equal to $0.05c$ and $\lambda_\parallel/r_g = 20$ (at maximum Jokipii limit). Note here that the spectrum is not changed while just two points are modified: 1) Getting to the maximum energy in a distribution of large and small jumps in momentum (see table 1); and 2) the speed of the acceleration process (obviously affected by the limits) with which we reach the spectrum. See figure 1. Bottom plot: Calculation of the differential time spectrum for a single run at $V_{sh}$ equal to $0.05c$ at $85^o$

Specifically, Wolf Rayet star (WR) winds could possibly provide large enough radii for the above conditions to be fulfilled, and on the other hand in Red Super Giant stars (RSG) winds we may find the conditions (smaller radii) where a cut off in energy is lower than the knee region. So, for RSG stars the maximum energy could be probably about $3 \times 10^{14}$ eV and we claim that for WR stars the maximum energy (for Z=1) reached could be around $10^{17}$ eV.

## 5. Summary and conclusions

In this work we tried to evaluate the rate of the maximum energy and the acceleration rate that particles acquire in the non-relativistic near perpendicular diffusive shock acceleration - allowing both parallel and perpendicular diffusion coefficients for the particle - with application to the stellar winds of Wolf Rayet (WR) and Red Super Giant (RSG) stars.

A Monte Carlo code was used to simulate the first order Fermi acceleration by testing the role between the obliquity of the magnetic field at the shock normal, the significance of the perpendicular (cross field diffusion) and parallel diffusion coefficients, and the energy gain and acceleration rate of the particles. Our findings justify Jokipii's work, who investigated the first order Fermi acceleration mechanism in connection to the high obliquity of the shock, and he showed that specific limitations should arise for the maximum energy gain of the particles connecting to the diffusion coefficients and the obliquity of the magnetic fields with the shock normal. Specifically if the perpendicular diffusion gets smaller compared to the parallel diffusion coefficient, the particles gain much more energy in few cycles (specifically, energies that could contribute to energies below the 'knee') than when the magnetic field at the shock is quasi-parallel. Provided that this condition applies, then we may claim the upper limits of the energy for sites where non-relativistic quasi-perpendicular shock velocities take place (e.g. $V_{sh} \leq 0.05c$).

Findings of this work are:

1. We confirm that the spectrum is independent of which scattering coefficient is used.
2. The energy gains are reached in large jumps, but only by a small percentage of the cosmic rays taking place in the simulation (less than 20% of the total number of cosmic rays used). Moreover, the individual steps to get to the spectrum are by a mixture of large and small steps in energy (not a gaussian distribution anymore, but more like a powerlaw distribution or a very asymmetric gaussian, with a long positive tail). In other words the distribution of $\frac{\Delta p}{p}$ has a long positive tail which lets us understand the faster acceleration rate while as it is well known, the spectrum remains unaffected by the obliquity of the shock.
3. The acceleration time scale is shorter, because the scattering is faster for the Jokipii approach. In other words in highly oblique shocks the energy jumps come both very large and very small, without changing the spectrum, but shortening the time scale of acceleration.
4. The cosmic rays reach very high energies (PeV) in a reasonable time limit frame using eq. (4) compared to standard Bohm limit ($r_g c/3$), see also figure (1). That means that highly oblique shocks help particles to get to 'ankle' (EeV) energies faster.

Summarising, our simulations have shown that in the highly oblique case we do reach effectively the Jokipii limit in the sense, that the effective acceleration time is given by equation (2), using Jokipii's limit. We recover the standard expression, and yet we learn that the standard derivation fails, in terms of the Bell (1978a,1978b) picture (with all the steps in momentum small). Jokipii's diffusion coefficient speeds acceleration up by a factor of $c/(3V_{sh})$ comparing his diffusion coefficient with the Bohm coefficient, where $c \sim u$ ($u$ is the particle speed). This matches what is found in our simulations as it is clear from equations (3) and (4). We further note that equation (2) is also valid in Jokipii's case, but has to be used with the coefficient parallel to the shock normal, which in the limit of a highly oblique shock is almost perpendicular to the magnetic field. A condition that it is quantitatively underpinned by Biermann (1993).

These simulations serve as an 'experiment' in order to investigate the shock-acceleration properties of the particles in the extreme condition of nearly perpendicular shocks. By measuring the 'simulation acceleration time' that a particle needs to get some energy including the parameters or limits of Jokipii, we can compare it with the standard expression. The diffusive shock acceleration mechanism is indeed efficient to accelerate particles to high energies. The more oblique the shock is the greater the efficiency of the shock to accelerate the particles.

It is quite plausible that a part of the spectrum just above the 'knee' and down close to the 'ankle' is likely due to the contribution of the cosmic rays, originating from regions where non-relativistic nearly perpendicular shocks take place, with a diffusion behaviour described in this work. The seed of particles to achieve such high energies may come from remnant winds of astrophysical sources. Our simulation findings underpin the past analytical work of Biermann (1993) and other connected works regarding cosmic ray acceleration in highly oblique shocks, which are mostly realised in magnetic stellar winds.

*Acknowledgements.* This work has been partially supported by the AUGER theory and membership grant 05 CU1ERA/3 via DESY/BMBF (Germany). The authors would like to thank Prof. J. Quenby and Prof. W. Rhode for helpful discussions and comments. AM would like to acknowledge the hospitality of the Physics groups of the University of Dortmund and Ruhr-University Bochum.